\documentclass[hyperlink,superscriptaddress,preprintnumbers,amsmath,amssymb,nofootinbib,letterpaper]{revtex4}
\usepackage{graphicx}
\usepackage{epstopdf}
\usepackage{dcolumn}
\usepackage{bm}
\usepackage{hyperref}
\usepackage{epsfig}
\def\slashchar#1{\setbox0=\hbox{$#1$} 
\dimen0=\wd0 
\setbox1=\hbox{/} \dimen1=\wd1 
\ifdim\dimen0>\dimen1 
\rlap{\hbox to \dimen0{\hfil/\hfil}} 
#1 
\else 
\rlap{\hbox to \dimen1{\hfil$#1$\hfil}} 
/ 
\fi}

\def\a{\alpha}

\def\c{\varepsilon}

\def\h{\theta}

\def\m{\mu}
\def\n{\nu}

\def\D{\Delta}
\def\G{\Gamma}

\def\beq{\begin{eqnarray}}
\def\eeq{\end{eqnarray}}

\newcommand{\lsim}{ \mathop{}_{\textstyle \sim}^{\textstyle <} }
\newcommand{\vev}[1]{ \left\langle {#1} \right\rangle }

\begin{document}
\title{Small steps towards Grand Unification \\ and \\ the 
 electron/positron excesses in cosmic-ray experiments
}
\author{Masahiro Ibe}
\affiliation{%
SLAC National Accelerator Laboratory, Menlo Park, CA 94025
}%


\begin{abstract}
We consider a small extension of the standard model
by adding two Majorana fermions; those are adjoint
representations of the $SU(2)_L$ and $SU(3)_c$ gauge groups
of the standard model.
In this extension, the gauge coupling unification at an energy scale higher than 
$10^{15}$\,GeV is realized when the masses of the triplet and the octet fermions
are smaller than $10^4$\,GeV and $10^{12}$\,GeV, respectively.
We also show that an appropriate symmetry ensures a long lifetime of the
neutral component of the triplet fermion whose thermal relic density
naturally explains the observed dark matter density.
The electron/positron excesses observed in recent cosmic-ray experiments 
can be also explained by the decay of the triplet fermion.
\end{abstract}

\date{\today}
\maketitle
\preprint{SLAC-PUB-13689}

\section{Introduction}
The most beautiful framework for physics beyond the standard model is 
the grand unified theory (GUT) where the three gauge groups of the model 
are unified into a larger gauge group at a very high energy scale\,\cite{Langacker:1980js}.
The most important prediction of the grand unified theory is 
the unification of the gauge coupling constants
at a very high energy scale.
Remarkably, the extrapolations of the three gauge coupling constants 
of the standard model to higher energies  roughly 
suggest the unification of the gauge coupling constants.

Despite such a suggestion of the coupling unification, however, the precise measurements
of the gauge coupling constants have
revealed that three couplings do not coincide
on one scale\,\cite{Ellis:1990wk}.
Another important prediction of the grand unified theory
is the finite lifetime of the proton which decays via the interactions mediated by
the heavy gauge bosons of the grand unified gauge theory.
Unfortunately, the predicted lifetime, $\tau(p\to \pi^0e^+)\sim 10^{30}$\,yr, 
is much shorter than the current experimental limit, $\tau(p\to \pi^0e^+)>8.2\times 10^{33}$\,yr\,\cite{:2009gd}.%
\footnote{Here, we are assuming the minimal gauge group of the grand unification, $SU(5)$,
where the leptons and quarks are classified into the $\bar{\bf 5}$
and ${\bf 10}$\,\cite{Langacker:1980js}.}

The above lessons tell us that the grand unified theory requires additional particles 
below the unification scale, 
so that the three gauge couplings  
better agree with each other on a high energy scale and the unification scale 
is high enough to suppress the rate of the proton decay. 
One of the most successful extension of the standard model which
satisfies those requirements
is the supersymmetric standard model where superpartners for all the standard model particles have masses 
of order of the electroweak scale\,\cite{Dimopoulos:1981yj}.
There, the unification is realized very precisely and the unification scale is raised 
to around $10^{16}$\,GeV, which predicts 
a much longer lifetime of the proton than the current experimental limit.

In this paper, we consider a much smaller extension of the standard model which
realizes the better unification and the higher unification scale
than those in the standard model.
Concretely, we just add two Majorana fermions; 
those are adjoint 
representations of the $SU(2)_L$ and $SU(3)_c$ gauge groups of the standard model.
We name them ``wino-like\,($\tilde w$)"  fermion and ``gluino-like\,($\tilde g$)" fermion, 
respectively, after the fashion of the supersymmetric standard model. 
The better unification and the higher unification scale are realized
when the masses of the adjoint fermions satisfy $M_{\tilde w}\lesssim 10^4$\,GeV
and $M_{\tilde g}\lesssim 10^{12}$\,GeV (see Ref.\,\cite{Krasnikov:1993sc} for an earlier
discussion on the effects of the adjoint fermions to the gauge coupling unification).

We  go one step further. 
The scale of the mass of the gluino-like fermion, 
$M_{\tilde g}\lesssim 10^{12}$\,GeV, is tempting to interrelate 
the mass to the breaking scale of the so-called Peccei--Quinn symmetry
which is introduced to solve the strong CP--problem\,\cite{Peccei:1977hh}.
As we will see, the mass hierarchy between the two adjoint fermions can be 
explained with an appropriate choice of the charges of the fermions
 under the Peccei--Quinn symmetry.

As an interesting bonus of the introduction of the Peccei--Quinn symmetry,
the interactions between the wino-like fermion and the 
fermions in the standard model can be suppressed.
As a result of the suppression, the neutral component of the wino-like fermion 
has a very long lifetime and is a candidate of the dark matter. 
In fact, the thermal relic density of the neutral wino-like fermion with a mass around $3$\,TeV
naturally explains the observed dark matter density, $\Omega_{\rm DM} h^2=0.1358^{+0.0037}_{-0.0036}$\,\cite{Komatsu:2008hk}.

As another bonus, 
the observed 
electron/positron excesses at the PAMELA\,\cite{Adriani:2008zr} and 
Fermi\,\cite{Abdo:2009zk} experiments are also
explained by the decay of the dark matter. 
As we will show, an appropriate charge assignment of the Peccei--Quinn symmetry
leads to a lifetime of the dark matter which is suitable to explain the 
electron/positron excesses in cosmic ray.

The organization of the paper is as follows.
In section\,\ref{sec:step1}, we show that the better unification and the higher unification scale
are realized when the masses of the adjoint fermions, $M_{\tilde w}$ and $M_{\tilde g}$,
satisfy $M_{\tilde w}\lesssim 10^4$\,GeV and $M_{\tilde g} \lesssim 10^{12}$\,GeV. 
In section\,\ref{sec:step2}, we consider a $U(1)$ symmetry to address
the origin of the mass hierarchy between the two adjoint fermions.
There, we show that the $U(1)$ symmetry can be identified with the Peccei--Quinn symmetry.
In section\,\ref{sec:step3}, we show that the neutral component of  the wino-like fermion
has a very long lifetime with an appropriate choice of the Peccei-Quinn charge 
assignment.
In section\,\ref{sec:excesses}, we demonstrate how well the excesses of the electron/positron 
fluxes observed at the PAMELA and Fermi experiments can be explained
by the decay of the wino-like fermion.
The final section is devoted to conclusions.

\section{Step 1: Coupling unification and masses of adjoint fermions}\label{sec:step1}

\begin{figure}[t]
 \begin{minipage}{.4\linewidth}
  \includegraphics[width=\linewidth]{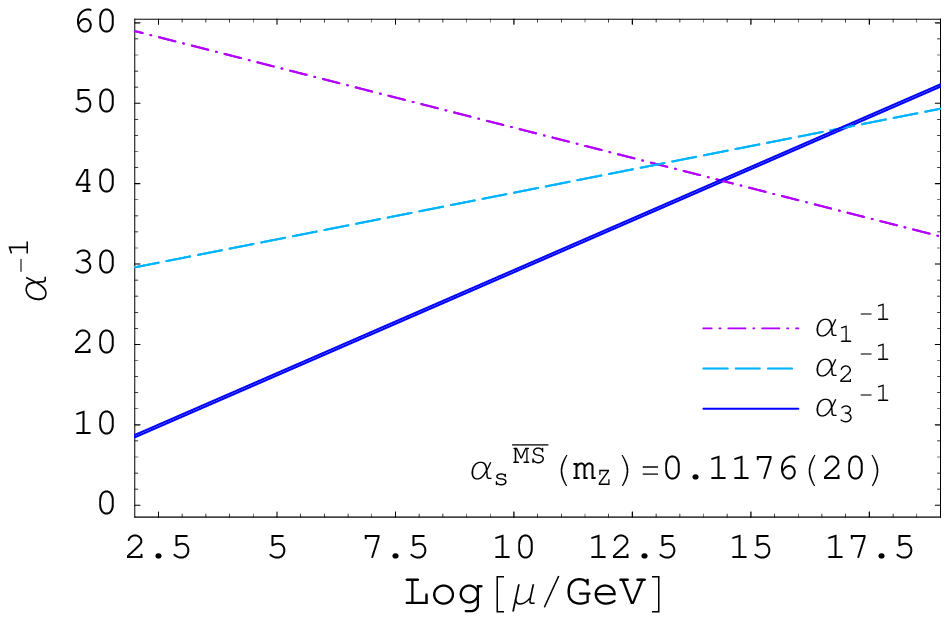}
 \end{minipage}
 \hspace{1cm}
 \begin{minipage}{.4\linewidth}
  \includegraphics[width=1.0\linewidth]{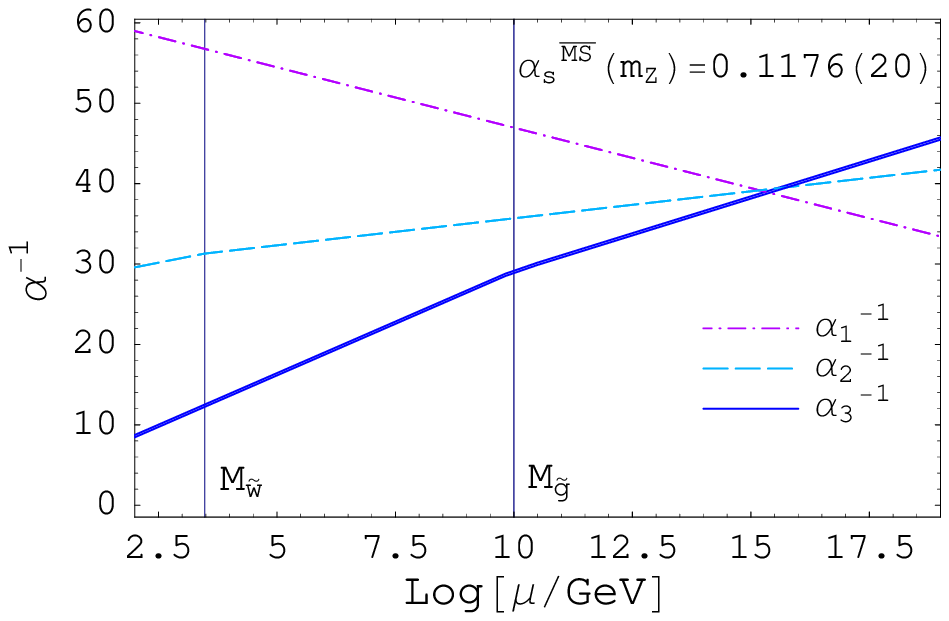}
 \end{minipage}
 \caption{The one-loop renormalization group evolutions of the gauge coupling constants in terms of $\a_a^{-1}$ in the standard model (left) and in the extended model (right).
  In  the extended model, the masses of the wino-like and gluino-like fermions
 are taken to be $M_{\tilde w}=3$\,TeV and $M_{\tilde g}=10^{10}$\,GeV, respectively.
 In the figures, we use $\alpha_3(m_Z)^{\overline{\rm MS}}=0.1176(20)$, $N_H =1$. 
We have also taken $m_{h}=117$\,GeV, and $m_{\rm top}=171.3$\,GeV,
although the results do not depend on these parameters significantly.
 }
\label{fig:CU}
\end{figure}

Let us begin with a basic test of the grand unification in an 
extension of the standard model with additional adjoint fermions;
the test on how well the three gauge coupling constants unify
at a high energy scale.
Throughout this paper, we assume the minimal gauge group of the grand unification, $SU(5)$,
where the leptons and quarks are classified into the $\bar{\bf 5}$ and $\bf 10$ representations\,\cite{Langacker:1980js}.
The unification scale is estimated by using the one-loop renormalization 
group equations of the gauge coupling constants,
\begin{eqnarray}
 \frac{d\a_a^{-1}}{d \ln \m} =- \frac{b_a}{2\pi}\,\,\, (a=1,2,3)\ ,
\end{eqnarray}
where $\m$ is the scale of the renormalization and the quantities $\alpha_i$ 
are related to the gauge coupling constants of the standard model gauge interactions
by $\alpha_i=g_i^2/4\pi$. Here, $g_1$ is a rescaled gauge coupling of the $U(1)_Y$
gauge interaction, {\it i.e.} $g_1 = \sqrt{5/3}g'$.
Above the electroweak scale, the coefficients of the beta functions are given by 
\begin{eqnarray}
 b_1 &=& 4+\frac{N_H}{10}\ ,\nonumber\\
  b_2 &=& \frac{10}{3}-\frac{N_H}{6} \,\quad(\m<M_{\tilde w}),
\quad 2-\frac{N_H}{6}\,\quad(\m>M_{\tilde w}),\nonumber\\
 b_3 &=& -7 \quad\quad\quad\quad\,(\m<M_{\tilde g}),\,\,\quad-5\quad\quad\quad\,\,(\m>M_{\tilde g}),
\end{eqnarray}
where $N_H$ is the number of Higgs doublets and $M_{\tilde w, \tilde g}$ denote the 
Majorana masses of the adjoint fermions.

In Fig.~\ref{fig:CU}, we compare the renormalization group evolutions of $\a_a^{-1}$
in the standard model with those in the extended model at the one-loop level.
In the extended model, we have taken $N_H = 1$, $M_{\tilde w}=3$\,TeV and 
$M_{\tilde g}=10^{10}$\,GeV as an example.
Contrary to the standard model, the gauge coupling constants in the extended model coincide
on one scale around $10^{15.5}$\,GeV.

\begin{figure}[t]
 \begin{minipage}{.35\linewidth}
  \includegraphics[width=\linewidth]{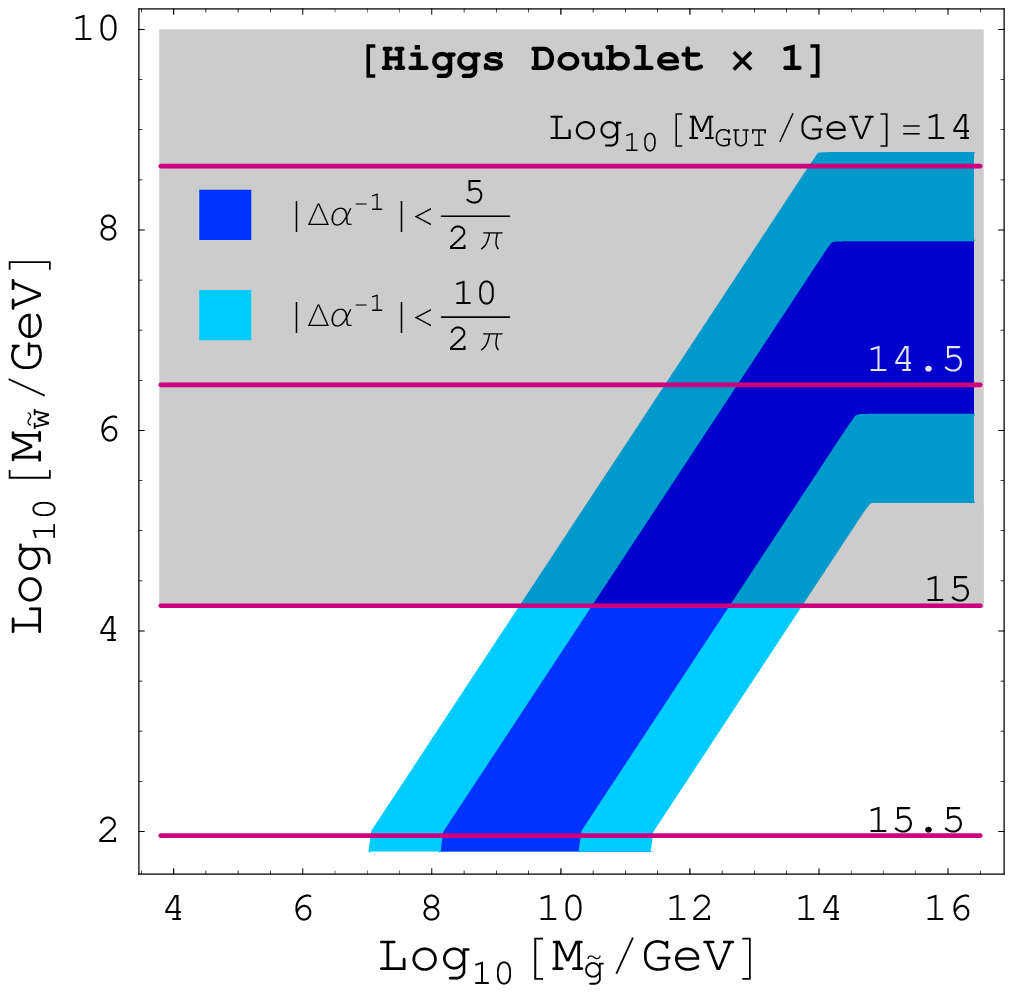}
 \end{minipage}
 \hspace{1.5cm}
  \begin{minipage}{.35\linewidth}
  \includegraphics[width=1.0\linewidth]{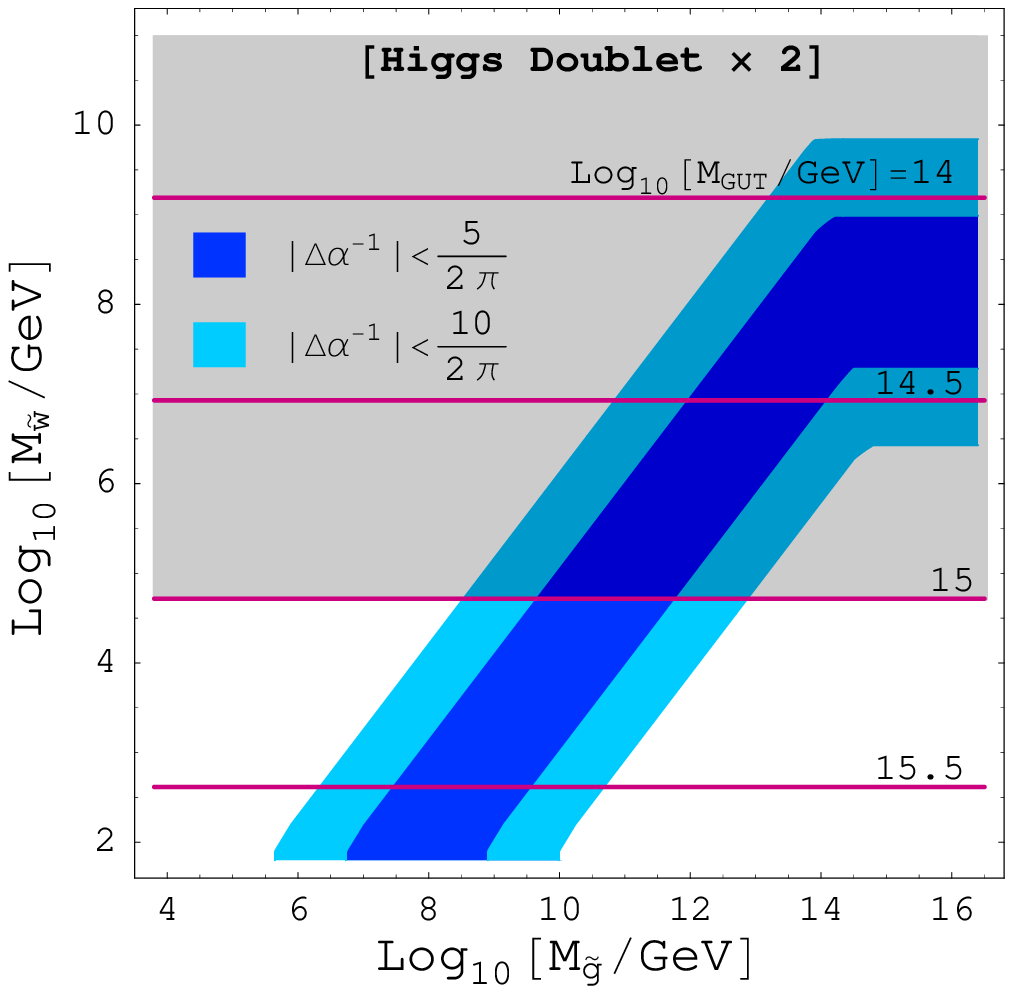}
 \end{minipage}
 \caption{
 The adjoint fermion masses which satisfy the unification test, {\it i.e.} $|N_{\rm th}| < 5, 10$
 (see Eq.\,(\ref{eq:NTH})), for one higgs doublet (left) and for two higgs doublets (right).
 The (light-)blue shaded regions correspond to the masses which satisfy $|N_{\rm th}|<5(10)$.
For $M_{\tilde g}\gtrsim 10^{15}$\,GeV, the degree of unification
does not depend on $M_{\tilde g}$ since $M_{\tilde g}$ is higher than the unification scale
 $M_{\rm GUT}$ in this region.
 The horizontal lines show the contours of the unification scale.
 The gray-shaded region corresponds to $M_{\rm GUT}<10^{15}$\,GeV
 which is roughly excluded by the current lower limit on the proton lifetime 
 (see discussion around Eq.\,(\ref{eq:proton})).
 In the figures, we have used $\alpha_3(m_Z)^{\overline{\rm MS}}=0.1176$,  
 $m_{h}=117$\,GeV, and $m_{\rm top}=171.3$\,GeV, although the results
 do not depend on those parameters significantly. 
 In the case of the two higgs doublet model, we assumed that the threshold corrections
 at the electroweak scale is not so different from those in the one higgs doublet model.
 This assumption is also good enough for our purpose  as long as the masses of the
  second higgs bosons are in the electroweak scale.
 }
\label{fig:unif}
\end{figure}

We quantify the degree of unification of the gauge coupling constants.
For that purpose, let us remind ourselves that there can be sizable 
threshold corrections to the gauge coupling constants
around the unification scale.
Therefore, the exact unification of the extrapolated gauge coupling constants does 
not have significant meaning, and there remains some freedom in how
we define the unification, which depends on explicit models of the grand unified theory.
In this study, instead of specifying models of the grand unified theory, we quantify
the degree of unification in terms of the size of the required threshold correction
at the unification scale, by defining the unification scale $M_{\rm GUT}$ and the threshold 
parameter $N_{\rm th}$ by,
\begin{eqnarray}
\label{eq:NTH}
\alpha_1(M_{\rm GUT} )= \alpha_2(M_{\rm GUT} )\equiv \alpha_{\rm GUT}\ ,\quad
{\mit\D} \alpha^{-1} = \alpha_{\rm GUT}^{-1} - \alpha_{3}^{-1}(M_{\rm GUT}) \equiv
\frac{N_{\rm th}}{2\pi}\ .
\end{eqnarray}
The parameter $N_{\rm th}$ quantifies how large a threshold correction at the unification scale
is required to realize a unified theory, and roughly speaking, it corresponds to 
the signed number of the charged particles (in the unit of the fundamental representation)
which contribute to the threshold correction around the unification scale.
For example, in the case of the supersymmetric standard model where 
masses of all the superparticles are of order of the electroweak scale, 
the threshold parameter satisfies $|N_{\rm th}|\lsim 5$\,\cite{Bagger:1995bw}.%
\footnote{The parameter $N_{\rm th}$ is related to the threshold parameter 
$\c_g$ in Ref.\,\cite{Bagger:1995bw}
by $\c_g = N_{\rm th}/4\pi\times\a_{\rm GUT} $.}

In Fig.\,\ref{fig:unif}, we show the degree of  unification 
in the $M_{\tilde g}$--$M_{\tilde w}$ plane for $N_H=1,2$.
The figures show that the precise unification, $N_{\rm th}\lesssim 5$, is realized for
\begin{eqnarray}
  M_{\tilde w}  \simeq 10^{-(6-8)}\times M_{\tilde g}\ ,\quad 
  {\rm or} \quad 
    M_{\tilde w}  \simeq 10^{6-8}\,{\rm GeV}\quad  {\rm for}\quad
 M_{\tilde g}  \gtrsim 10^{15}\,{\rm GeV}\ ,   
\end{eqnarray}
for $N_H=1$, and for
\begin{eqnarray}
  M_{\tilde w}  \simeq 10^{-(5-7)}\times M_{\tilde g}\ ,\quad
    {\rm or} \quad 
    M_{\tilde w}  \simeq 10^{7-9}\,{\rm GeV}  \quad  {\rm for}\quad
 M_{\tilde g}  \gtrsim 10^{15}\,{\rm GeV}\ ,   
\end{eqnarray}
for $N_H=2$.

Next, we consider the second test, the lifetime of the proton.
In the minimal grand unified theory with the $SU(5)$ gauge group,
the protons decay into  pairs of the pion and the electron 
via the effective four fermi interactions (see for example Ref.\,\cite{Ellis:1980jm}),
\begin{eqnarray}
{\cal L} = \frac{g_{\rm GUT}^2}{M_V^2}\left[ 
A_R\,(\bar{d}_R^{\dagger}\bar{u}_R^{\dagger})
(u_Le_L)
+
A_L(1+|V_{ud}|^2)
\,(u_Ld_L)(\bar{u}_R^{\dagger}\bar{e}_R^{\dagger}) + h.c.
\right]\ ,
\end{eqnarray}
which are mediated by the exchanges of the heavy gauge bosons of the grand unified theory.
Here, $g_{\rm GUT}$ is the unified gauge coupling constant, 
$g_{\rm GUT}^2/4\pi \simeq 1/40$, $M_V$ the mass of the heavy gauge bosons,
$V_{ud}\simeq 0.974$ the $ud$-component of the Cabibbo-Kobayashi-Masukawa matrix.
The coefficients  $A_{R,L}$ represent the renormalization factors of the above operators
from the unification scale to the lower energy scales.
At the renormalization scale $\m=2$\,GeV, the coefficients $A_{R, L}$ 
are given by,
\begin{eqnarray}
 A_{R,L} &=& A_{R,L}^{\rm SM}
 \times 
 \left( \frac{\a_2(M_{\tilde w})}{\a_{\rm GUT}} \right)^{\frac{27}{12}\left(
b_{2}^{-1}(\m>M_{\tilde w})
-
b_{2}^{-1}(\m<M_{\tilde w})
\right)
  }
  \times 
 \left( \frac{\a_3(M_{\tilde g})}{\a_{\rm GUT}} \right)^{2\left(
b_{3}^{-1}(\m>M_{\tilde g})
-
b_{3}^{-1}(\m<M_{\tilde g})
\right)
}\ ,\cr
&\simeq& A_{R,L}^{\rm SM}\times(1.0-1.2)\ ,
\end{eqnarray}
for wide ranges of $M_{\tilde w}$ and  $M_{\tilde g}$.
The renormalization factors in the standard model, $A_{R,L}^{\rm SM}$,
are given by 
$A_R^{\rm SM} \simeq 3$ and $A_L^{\rm SM} \simeq 3.2$
at $\m=2$\,GeV\,\cite{Ellis:1980jm}.%
\footnote{The renormalization factors are slightly smaller than those in Ref.\,\cite{Ellis:1980jm}
due mainly to the use of the different standard model parameters.}
From the above operators, the lifetime of the proton is given by,
\begin{eqnarray}
\tau(p\to \pi^0e^+)\simeq 1.4\times 10^{34}\,{\rm yr}
\times
\left(\frac{A_{L,R}^{\rm SM}}{A_{L,R}}\right)^2
\left(\frac{1/40}{\a_G}\right)^2
\left(\frac{M_V}{10^{15.5}\,\rm GeV}\right)^4
\left(\frac{0.06\,{\rm GeV}^2}{|W_0|}\right)^2\ ,
\end{eqnarray}
where $W_0=-0.06\pm 0.018$\,GeV$^2$ is the form factor of the proton decay operators
between the proton and the pion states calculated with lattice QCD\,\cite{Aoki:2006ib}.%
\footnote{The proton decay rate with $W_0=-0.06$\,GeV$^2$
corresponds to that expressed in terms of the form factor with the proton and the vacuum, $\a_H=0.005$\,GeV$^3$,
which is often used to represent the proton lifetime in the literature.
In the chiral perturbation theory, those parameters are related by $W_0=\a_H(1+g_A)/\sqrt{2}f_\pi$,
with the tree-level pion decay constant $f_\pi \simeq 131$\,MeV and the nucleon axial charge, $g_A\simeq 1.22$.
See Ref.\,\cite{Aoki:2006ib} for detailed discussions on the lattice simulations 
on those form factors.
}

By comparing the predicted lifetime with the current experimental limit, $\tau(p\to \pi^0e^+)>8.2\times 10^{33}$\,yr\,\cite{:2009gd}, we obtain a lower limit on the heavy gauge boson mass,
\begin{eqnarray}
 M_V \gtrsim 10^{15.4}\,{\rm GeV}\times
\left(\frac{A_{L,R}}{A_{L,R}^{\rm SM}}\right)^{1/2}
\left(\frac{\a_G}{1/40}\right)^{1/2}
\left(\frac{|W_0|}{0.06\,{\rm GeV}^2}\right)^{1/2}\ .
\end{eqnarray}
Then, by expecting that the mass of the heavy gauge bosons is not so far from the
unification scale, we can translate the above lower limit to a limit on the unification scale.
Notice that the exact relation between the unification scale and the gauge boson mass
depends on models of the grand unified theory.
In this study, instead of specifying models of the unified theory, 
we just assume that the gauge boson mass is of order of the unification scale
and we put a rough lower limit on the unification scale,
\begin{eqnarray}
\label{eq:proton}
 M_{\rm GUT}\gtrsim 10^{15}\,{\rm GeV}\ .
\end{eqnarray}

In Fig.\,\ref{fig:unif}, 
the shaded regions satisfy 
the test of the proton lifetime; $M_{\rm GUT}\gtrsim 10^{15}$\,GeV.
The figures show that the regions of relatively heavy wino-like fermion 
are excluded by the second test.
As a result, the masses of the adjoint fermions which pass both the tests are as follows;
 \begin{eqnarray}
  M_{\tilde w}  \simeq 10^{-(6-8)}\times M_{\tilde g}\quad \,\,
( M_{\tilde w}  \lesssim 10^{4}\,{\rm GeV} ),   
\end{eqnarray}
for $N_H=1$ and
\begin{eqnarray}
  M_{\tilde w}  \simeq 10^{-(5-7)}\times M_{\tilde g}\,\,\,\,
( M_{\tilde w}  \lesssim 10^{4.5}\,{\rm GeV} ),   
\end{eqnarray}
for $N_H=2$.

\section{Step 2: Origin of masses of adjoint fermions}\label{sec:step2}
In the previous section, we have shown that the small extension
of the standard model with adjoint fermions predicts 
better unification with a longer lifetime of the proton
than those in the standard model for
\begin{eqnarray}
M_{\tilde w} \lesssim 10^{4}\,{\rm GeV}\ ,\quad  M_{\tilde g} \lesssim 10^{12}\,{\rm GeV}\ .
\end{eqnarray}
In this section, we try to explain the mass spectrum of the adjoint fermions
by considering a spontaneous symmetry breaking of a global $U(1)$ symmetry.
As we will see, the $U(1)$ symmetry can be identified with the Peccei--Quinn 
symmetry\,\cite{Peccei:1977hh}, 
and hence, the strong CP--problem is solved automatically.

Let us first assume that the model is invariant under global $U(1)$ chiral rotations,
\begin{eqnarray}
 \tilde{g} \to \tilde{g}' = e^{i\a/2} \tilde{g}\ ,\quad
  \tilde{w} \to \tilde{w}' = e^{i\a} \tilde{w}\ ,
\end{eqnarray}
with an angle $\a$.%
\footnote{The above charge assignment suggests that the 
gluino-like and the wino-like fermions stem from
different multiplets in the grand unified theory,
although we do not pursue explicit models of the grand unified theory in this paper.}
Under this symmetry, the masses of the adjoint fermions are forbidden.
Next let us further assume that the chiral symmetry is broken spontaneously
at around $10^{8-12}$\,GeV
by a condensation of a scalar field $X$ which rotates under the above chiral symmetry by,
\begin{eqnarray}
 X\to X' = e^{-i\a}X\quad  (\vev{X} \simeq 10^{8-12}\,{\rm GeV}).
\end{eqnarray}
With this spontaneous breaking, the gluino-like fermion obtains a mass 
from a direct coupling with $X$,
\begin{eqnarray}
 {\cal L}_{\tilde g} \simeq \frac{1}{2}X\, \tilde{g} \tilde{g} + h.c.\ ,
\end{eqnarray}
which results in $M_{\tilde g}\simeq \vev {X}\simeq 10^{8-12}$\,GeV. 
Here, we have neglected coefficients of the order one.

The mass term of the wino-like fermion, on the other hand, is still suppressed by the chiral symmetry,
and it begins with a dimension five operator suppressed by $M_{\rm GUT}$,
\begin{eqnarray}
  {\cal L}_{\tilde w} \simeq \frac{1}{2}\frac{X^2}{M_{\rm GUT}}\, \tilde{w} \tilde{w} + h.c.\ ,
\end{eqnarray}
when we assume that the interactions between the scalar $X$ and the wino-like fermion
are mediated by fields of masses of order of the unification scale.
Once the dimension five operator is generated at the unification scale, 
the mass of the wino-like fermion is given by,
\begin{eqnarray}
 M_{\tilde w} \simeq \frac{\vev{X}^2}{M_{\rm GUT}} \simeq 3\,{\rm TeV} 
\times \left(\frac{\vev X}{10^{9.5}\,\rm GeV}\right)^2
 \left(\frac{10^{15.5}\,\rm GeV}{M_{\rm GUT}}\right)\ ,
\end{eqnarray}
which is consistent with the unification tests in the previous section.
As a result, we found that the masses of the adjoint fermions 
can be naturally explained by spontaneous breaking of a chiral symmetry 
at the intermediate scale.

The interesting outcome of the above chiral symmetry is that the chiral symmetry 
plays the role of the so-called Peccei--Quinn symmetry\,\cite{Kim:1979if,Zhitnitsky:1980tq}.
That is, the above chiral rotation is anomalous to the $SU(3)_c$
gauge symmetry, and the axion resulting from the spontaneous breaking of the 
chiral symmetry cancels the $\theta$ angle in QCD which is otherwise
required to be tuned to a very small value, $|\theta| \lsim 10^{-10}$\,\cite{Baker:2006ts}.
It should be noted that the properties of the axion 
are consistent with the astrophysical and the cosmological constraints for
$\vev{X}\simeq10^{9-12}$\,GeV
(see for example Ref.\,\cite{Raffelt:2006cw}
and references there in for detailed discussion on astrophysical constraints 
on the Peccei--Quinn breaking scale).
Therefore, the above small extension of the standard model which satisfies 
the unification can be naturally integrated with the solution to the strong CP--problem.

\section{Step 3: Stability of adjoint fermions and dark matter density}\label{sec:step3}
Recent observations of the electron/positron excesses in
the PAMELA\,\cite{Adriani:2008zr}
and Fermi\,\cite{Abdo:2009zk} experiments strongly suggest the existence of a new source of  electron/positron fluxes.  
The most interesting candidate of the new source 
which is related to physics beyond the standard model is
the decay of the dark matter with a mass in the TeV range. 
Therefore, it is an interesting question whether the above wino-like fermion can 
be a candidate of the dark matter, and on top of that, it explains the observed 
electron/psoitron excesses in cosmic ray.
(The earlier works on the electron/positron excesses 
from the decay of the wino-like dark matter of a mass in the TeV
range have been done in Refs.\,\cite{Nardi:2008ix,Ibarra:2008jk} 
in a model independent way,
and in Refs.\,\cite{Shirai:2009fq,Chen:2009mj} 
in the context of the supersymmetric standard model.)

Before going to the stability of the wino-like fermion, it should be checked whether
the neutral component of the triplet wino-like fermion
is the lightest component.
The dominant mass splitting between the neutral and the charged components
in the wino-like fermion comes from the one-loop weak gauge boson exchange 
diagrams~\cite{Feng:1999fu}, which is given by
\begin{eqnarray}
\label{eq:splitting}
{\mit \D} M_{\tilde w} &=&  M_{{\tilde w}^{\pm}}-M_{{\tilde w}^{0}}
=\frac{g_{2}^{2}}{16\pi^{2}} M_{\tilde{w}} 
\left[ f(r_{W})-\cos^{2}\h_{W} f(r_{Z})-\sin^{2}\h_{W} f(0) \right],\nonumber\\
&\simeq& 161\,{\rm MeV}-165\,{\rm MeV}
\quad ( {\rm for}\,\,M_{\tilde w}=1\,{\rm TeV}-10\,{\rm TeV}),
\end{eqnarray}
where $f(r)= \int^{1}_{0} dx(2 + 2 x^{2}) \ln[x^{2} + (1-x)r^{2}]$, 
$r_i$ denotes the weak gauge boson masses normalized by the mass of wino-like fermion, 
$r_{i}=m_{i}/M_{\tilde w}$. 
Here, we have assumed that the direct interactions between the wino-like
fermion and the standard model fields are suppressed, which will be 
justified in the following discussion.
As a result, we see that the neutral component is the lightest component of the 
wino like fermion.
This splitting allows the charged components decay into the neutral component 
and a virtual $W^{\pm}$ bosons which end up with $\pi^{\pm}$ or lepton pairs.
These are crucial features of the wino-like fermion as a dark matter candidate,
otherwise the wino-like fermion leads to a charged dark matter.

Now, let us ask whether the neutral component is stable or not.
The easiest way to achieve the stability is to introduce
a $Z_2$ symmetry under which the wino-like fermion changes the sign.
With the $Z_2$ symmetry, 
we can forbid any interactions which cause the decay of the neutral component of 
the wino-like fermion.
It is, however, more attractive if the stability of the neutral component 
is ensured by  symmetries which are introduced for some other reasons than the
the stability of the dark matter.
In the followings, we show that an appropriate charge assignment of the Peccei--Quinn symmetry
to the standard model fields leads to a stability of the wino-like fermion.

The lowest dimensional interactions which cause  the decay of the wino-like fermion are given by,
\begin{eqnarray}
\label{eq:decay1}
  {\cal L}_{\rm decay} = c_i H^*_{a}\tilde{w}^A t^{Aa}_b \ell_{Li}^b\ , 
\end{eqnarray}
where $c_i$ denotes a coefficient, $t^A$ the generators of $SU(2)$,
$H$ the higgs doublet, and $\ell_{Li}$  the lepton doublets of the flavor indices $i=1,2,3$.
In the followings, we assume $N_H=1$, although we can extend our discussion for $N_H=2$, straightforwardly.
With the above operators, the decay rate of the wino-like fermion is given by,
\begin{eqnarray}
\label{eq:decay}
 \G_{\tilde w}(\tilde{w}^0\to \ell^{\pm}W^{\mp},\n Z, \n h)\simeq\, \frac{c_i^2}{2\pi}M_{\tilde w}\ .
\end{eqnarray}
Here, we have summed all the possible final states.
Therefore, in order for the wino-like fermion to be a dark matter candidate,
$c_i$ must be highly suppressed. 

\begin{table}[tb]
\begin{center}
\begin{tabular}{c|ccccccc}
    &$H$ & $\ell_{L}$ & $\bar{e}_{R}$ & $q_{L}$ & $\bar{u}_{R}$ & $\bar{d}_{R}$ & $\tilde{w}$ \\
    \hline
$SU(2)$ &${\bf 2}$ &${\bf 2}$ &${\bf 1}$ & ${\bf 2}$ & ${\bf 1}$ & ${\bf 1}$ & ${\bf 3}$\\
$U(1)_{\rm Y}$ & $-1/2$  & $-1/2$ &$1$ &  $1/6$ &$-2/3$ &  $1/3$ & $0$ \\
 $U(1)_{\rm PQ}$& $12/5$& $-18/5$ & $6/5$ & $6/5$ & $6/5$  & $-18/5$ & $1$
\end{tabular}
\end{center}
\caption{The Peccei--Quinn-charges of the standard model fields.
Here, we also show the charges under the standard model gauge group; 
$SU(2)\times U(1)_{Y}$. 
We assign the same charges to all the three generations of the standard model fermions. }
\label{tab:charge}
\end{table}%

To suppress the above operators, 
we give charges to the standard model fields under the Peccei--Quinn symmetry.
In Table.\,\ref{tab:charge}, we give an example of the charge assignment
which suppresses the operators in Eq.\,(\ref{eq:decay1}).
With this charge assignment, the operators in Eq.\,(\ref{eq:decay1}) 
have the corresponding Peccei--Quinn charge $-5$, and hence, they are highly suppressed.
That is, when the interactions between the scalar $X$ and the standard model fields as well as
the wino-like fermions are mediated by the fields of masses of order of the unification scale,
the above operators only come from the effective operators,
\begin{eqnarray}
{\cal L}_{\rm decay} \simeq \frac{X^{*5}}{M_{\rm GUT}^5}H^*_{a}\tilde{w}^A t^{Aa}_b L^b_i\ , 
\end{eqnarray}
and hence, the coefficient $c_i$ is highly suppressed by $(\vev{X}/M_{\rm GUT})^5$.
As a result of the suppression, the wino-like fermions 
decay and the lifetime of the neutral wino-like fermion is given by,
\begin{eqnarray}
 \tau_{\tilde w} \simeq 4\times 10^{25}\,{\rm sec}\times
  \left(
 \frac{3\,{\rm TeV}}{M_{\tilde w}}
 \right)
 \left(
 \frac{10^{9.5}\,{\rm GeV}}{\vev{X}}
 \right)^{10}
 \left(
 \frac{M_{\rm GUT}}{10^{15.5}\,{\rm GeV}}
 \right)^{10}\ .
\end{eqnarray}
Therefore, the lifetime of the neutral wino-like fermion can be long enough to be a dark matter candidate.
Moreover, the lifetime is in an appropriate range to explain the electron/positron excesses
by the decay of the dark matter.%
\footnote{
The other higher dimensional operators which also cause the decay of the neutral wino-like fermion are more suppressed by factors of the unification scale and by the symmetry. 
It should be also noted that when there are  four fermion interactions such as 
$\tilde {w} \bar{e}_R \ell_{Li}\ell_{Lj}$, the operators in Eq.\,(\ref{eq:decay1})
are induced radiatively and dominate the decay process.
Thus, in the non-supersymmetric models, the decay modes via the four fermion
interactions such as $\tilde w \to \ell+\ell+\n$  
cannot be the dominant mode.
}

\begin{figure}[t]
 \begin{minipage}{.35\linewidth}
  \includegraphics[width=\linewidth]{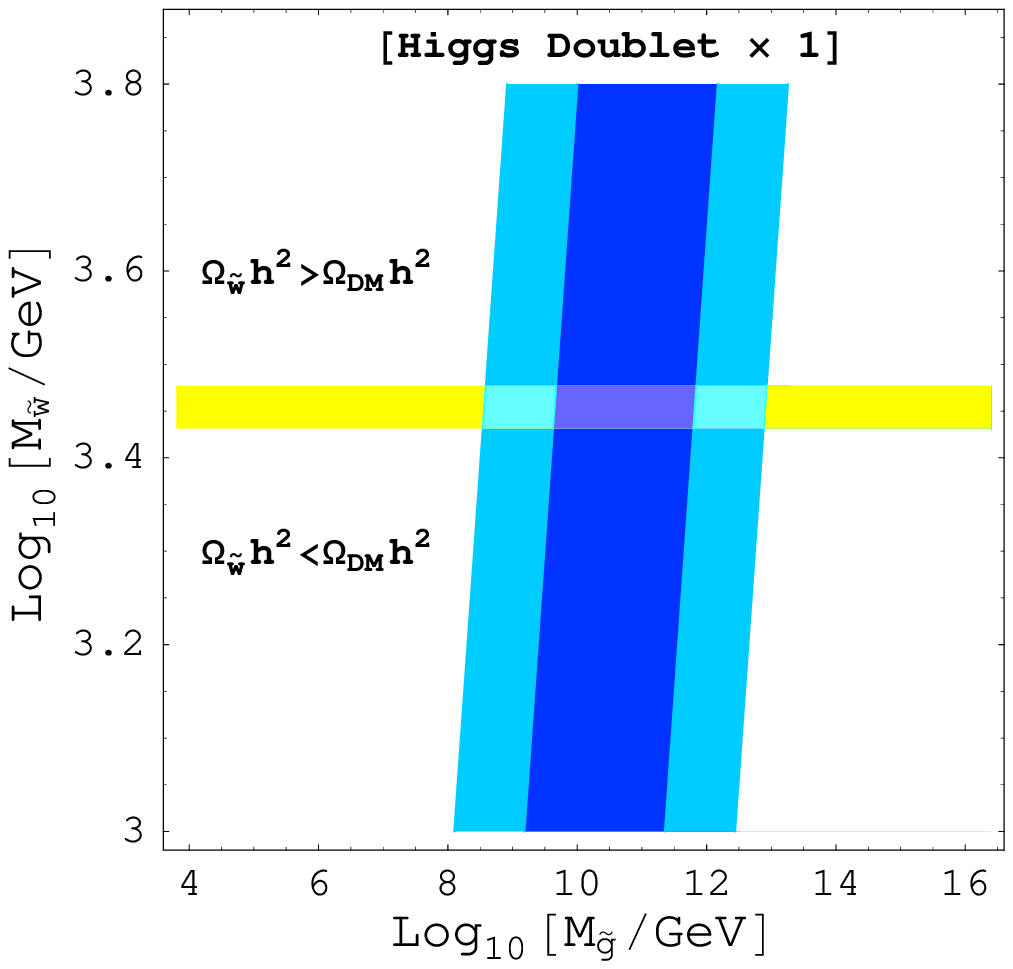}
 \end{minipage}
 \hspace{1.5cm}
  \begin{minipage}{.35\linewidth}
  \includegraphics[width=1.0\linewidth]{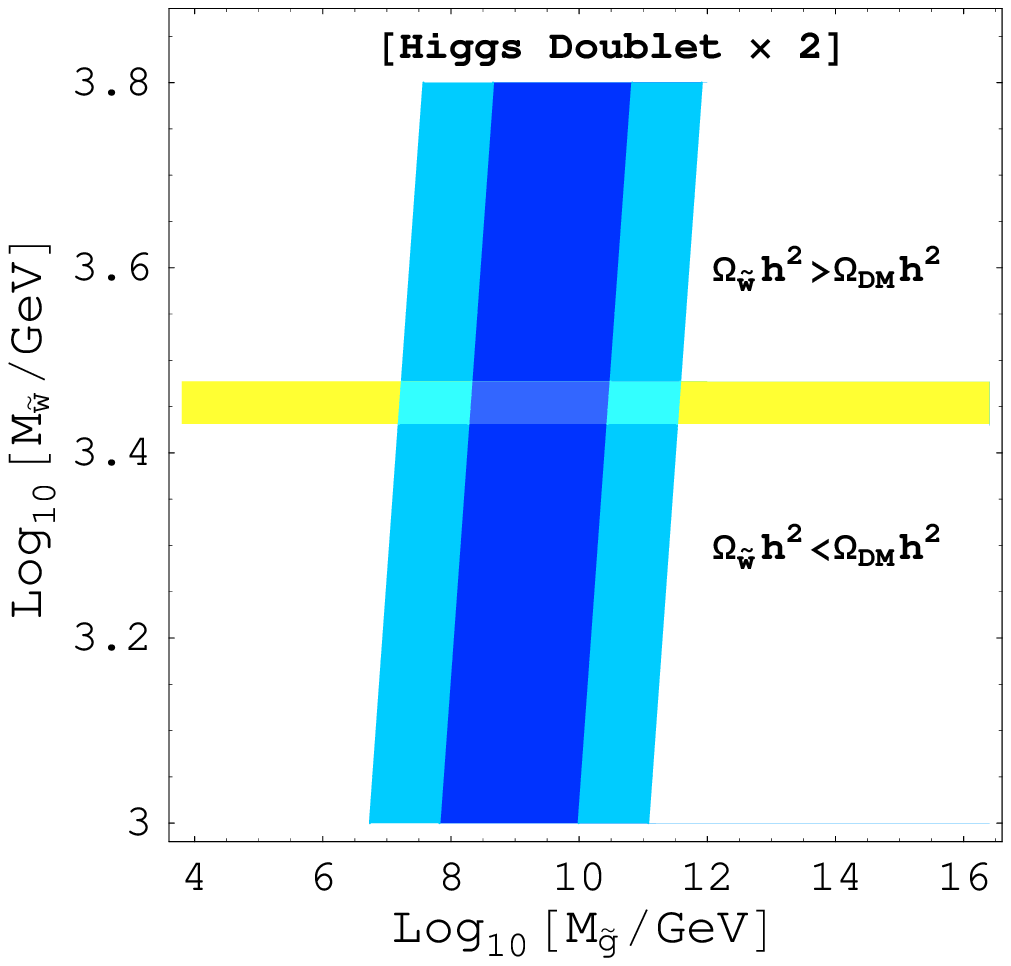}
 \end{minipage}
 \caption{
 The enlarged views of Fig.\,\ref{fig:unif}.
 On the yellow bands, the thermal relic density is consistent with the observed dark matter density,
 which corresponds to $2.7$\,TeV$\lesssim M_{\tilde w}\lesssim 3$\,TeV.
 Above the bands, the thermal relic density is larger than the observed dark matter density,
 while it is smaller below the bands.
 }
\label{fig:unif_dm}
\end{figure}

We next consider the thermal relic abundance of the wino-like fermion.
Since the interactions between the wino-like fermion and the matter fields of the standard model 
are highly suppressed, the dominant annihilation process is the one into two $W$-bosons via the $t$-channel
exchange of the charged wino-like fermions.
The perturbative analysis on this process gives the thermal relic density of the wino-like fermion
\cite{Fujii:2003nr},
\begin{eqnarray}
\Omega_{\tilde w}h^2 \simeq 0.1\times 
\left(\frac{M_{\tilde{w}}}{2\,{\rm TeV}}\right)^2\quad ({\rm perturbative}).
\end{eqnarray}
As pointed out in Ref.\,\cite{Hisano:2006nn}, however, the thermal relic abundance 
is significantly changed by a non-perturbative effect called 
the Sommerfeld enhancement when the neutral and the charged wino-like fermions are almost degenerate.
The resultant mass range of the wino-like fermion which is consistent with the observed
dark matter density is then given by\,\cite{Hisano:2006nn},
\begin{eqnarray}
 2.7\,{\rm TeV}\lesssim M_{\tilde w} \lesssim 3.0\,{\rm TeV}\quad(\mbox{non-perturbative}).
\end{eqnarray}
In Fig.\,\ref{fig:unif_dm}, we show the corresponding parameter region
on the $M_{\tilde g}-M_{\tilde w}$ plane.
From the figure, we see that the masses  which have passed the unification tests
are consistent with the mass of the wino-like fermion
which explain the observed dark matter density.

Before closing this section, we comment on the contributions of the 
standard model quarks to the $SU(3)_c$ anomaly of the Peccei-Quinn symmetry.
As we see from the Table\,\ref{tab:charge}, the contribution in the quark sector 
is cancelled, {\it i.e.}
\begin{eqnarray}
 \frac{6}{5}(2+1)-\frac{18}{5}=0\ ,
\end{eqnarray}
where the first term denotes the contribution from $q_L$ and $\bar{u}_R$, 
while the second term from $\bar{d}_R$.
This cancellation can be understood by remembering that the above charge
assignment can be expressed by,
\begin{eqnarray}
Q_{\rm PQ}^{\rm SM} = -\frac{24}{5}Q_Y -6 Q_{B-L}\ ,
\end{eqnarray}
where both of $Q_Y$ and $Q_{B-L}$ are anomaly free.
Therefore, the anomaly of the Peccei--Quinn symmetry to the
$SU(3)_c$ gauge symmetry only comes from the gluino-like fermion
under this charge assignment.

\section{Electron/positoron excesses from decay of dark matter}\label{sec:excesses}

\begin{figure}[t]
 \begin{minipage}{.38\linewidth}
  \includegraphics[width=\linewidth]{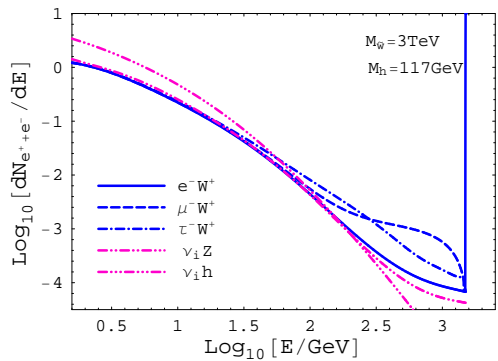}
 \end{minipage}
 \caption{
 The fragmentation functions into electrons and positrons of the each decay modes. 
 We used the program PHYTIA\,\cite{Sjostrand:2006za} to obtain  the functions.
 In this figure, we have taken $M_{\tilde w}=3$\,TeV and $m_h = 117$\,GeV.
  }
\label{fig:specSM}
\end{figure}
In the previous section, we showed that 
the wino-like fermion is a good candidate of the dark matter 
with the help of the Peccei--Quinn symmetry,
and the thermal relic density explains the observed dark matter density
for $M_{\tilde w}\simeq 3$\,TeV.
Besides, with an appropriate choice of the Peccei--Quinn charges,
the lifetime of the dark matter can be in an appropriate range for an explanation 
of the observed electron/positron excesses in cosmic ray, {\it i.e.} $\tau_{\tilde w}\sim 10^{26}$\,sec.
In fact, the decay mode of the wino-like fermion, 
$\tilde{w}\to \ell^\pm W^\mp $ has been studied extensively
in Ref.\,\cite{Nardi:2008ix,Ibarra:2008jk}, and the observed cosmic-ray spectra can be well fitted
for $\ell = \mu$ with $M_{\tilde w}=3$\,TeV.

In this section, we apply their analysis to the wino-like fermion dark matter
which decays via the operators in Eq.\,(\ref{eq:decay}).
According to the results in Refs.\,\cite{Nardi:2008ix,Ibarra:2008jk}, 
we concentrate on the case with $|c_2|\gg|c_{1,3}|$, so that 
the mode into $\ell = \mu$ is the dominant one.
The important difference of our analysis from the generic analysis 
is that the decay mode $\tilde{w}\to \ell^{\pm}W^{\mp}$ is 
accompanied by the other decay modes, $\tilde{w}\to \n Z$ and $\tilde{w}\to \n h$ 
with the branching ratios,%
\footnote{
The above branching ratios are similar to the ones
considered in the decaying gravitino dark matter scenarios\,\cite{Ishiwata:2009vx,
Buchmuller:2009xv}.
}
\begin{eqnarray}
\label{eq:branching}
{\rm Br}(\tilde w \to \m^{\pm}W^{\mp}) =0.5,\quad
{\rm Br}(\tilde w \to \n_2^{(\dagger)} Z) =
{\rm Br}(\tilde w \to \n_2^{(\dagger)}h)  = 0.25.
\end{eqnarray}
In Fig.\,\ref{fig:specSM}, we show the fragmentation functions of the each decay modes
into electrons and positions for $M_{\tilde w}=3$\,TeV.
The figure shows that the contribution from the $Z$ and $h$ modes 
increase the number of the low energy electrons and positirons compared 
with those in the pure $W$ mode.

\begin{figure}[htb]
 \begin{minipage}{.35\linewidth}
  \includegraphics[width=\linewidth]{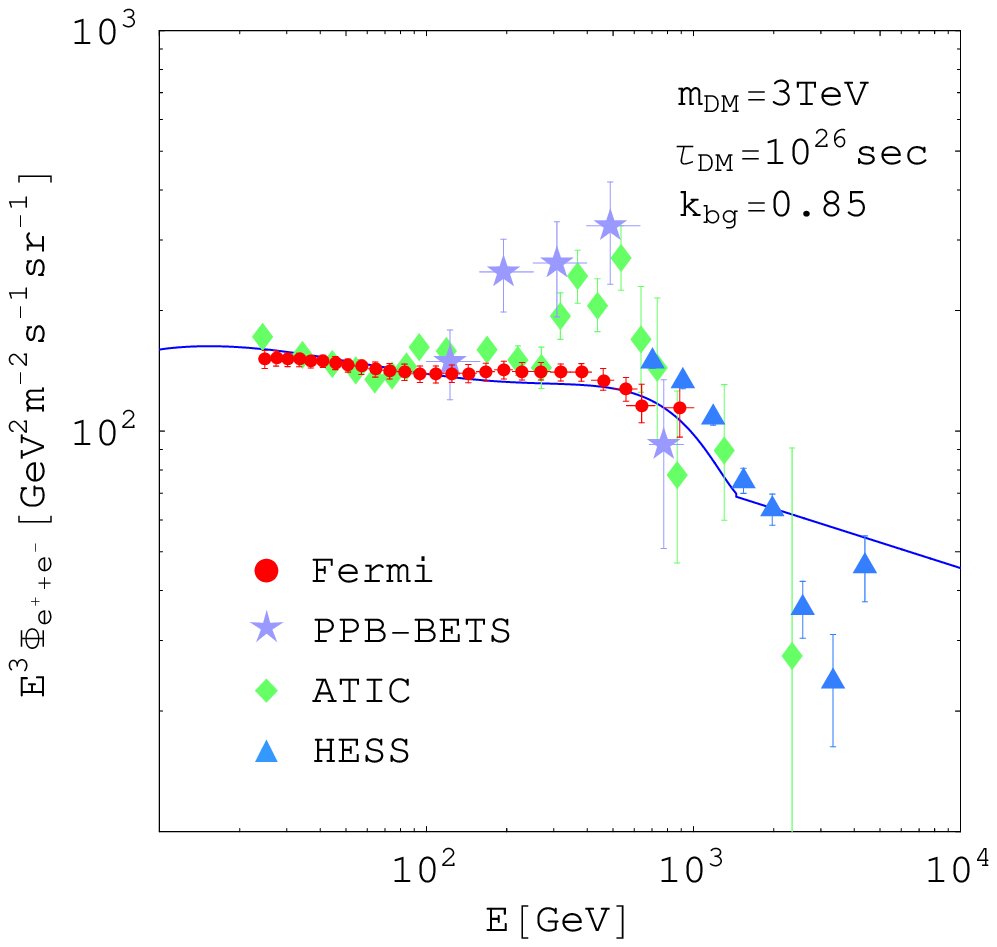}
 \end{minipage}
 \hspace{1.5cm}
  \begin{minipage}{.3715\linewidth}
  \vspace{.25cm}
  \includegraphics[width=1.0\linewidth]{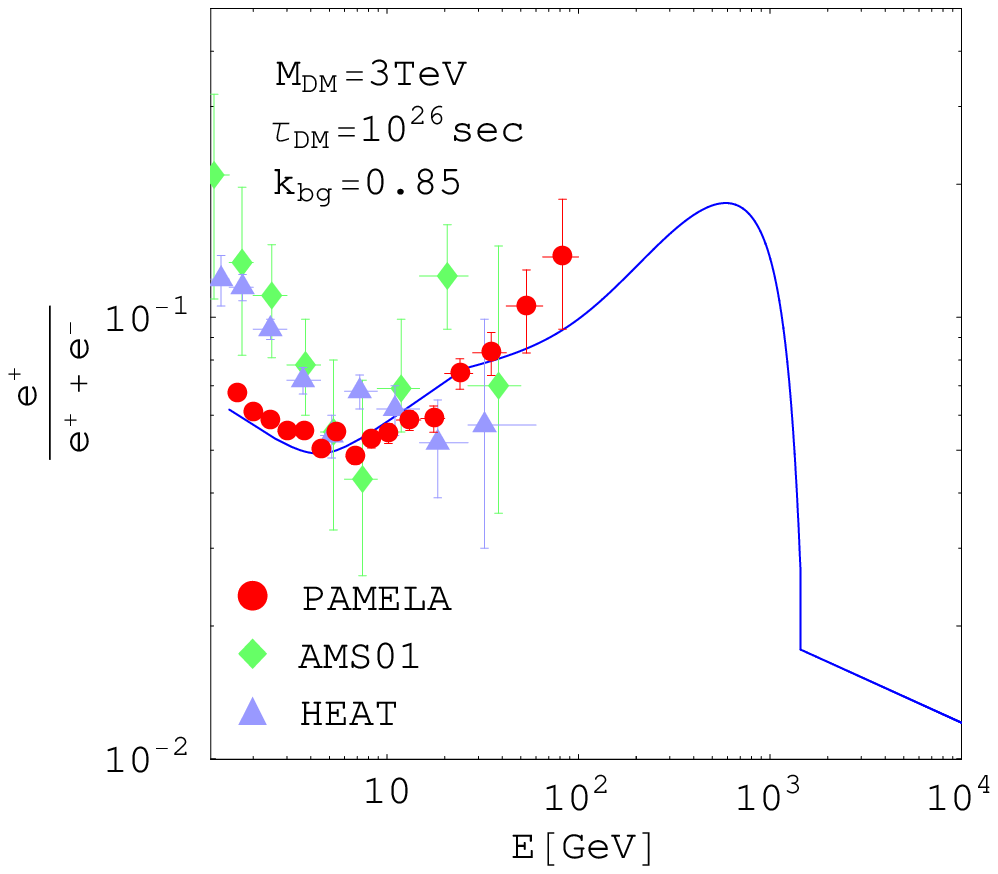}
 \end{minipage}
 \caption{
 Left) The predicted electron/positron flux in cosmic ray for the branching ratio in 
 Eq.\,(\ref{eq:branching}).
 The propagation of the electron/positron flux in the galaxy is analyzed based 
 on a numerically approximated Green function given in Ref.\,\cite{Ibarra:2008qg}
(the MED propagation model).
The prediction is compared with the experimental 
data\,\cite{Abdo:2009zk,Chang:2008zz,Torii:2008xu,Collaboration:2008aaa}.
Right) The predicted positron fraction in  cosmic ray for the same dark matter decay modes
with the experimental data\,\cite{Adriani:2008zr,Aguilar:2007yf,Barwick:1997ig}.
 }
\label{fig:electron}
\end{figure}

The predicted electron/positron spectrum in cosmic ray 
is shown in Fig.\,\ref{fig:electron}
for the dark matter lifetime $\tau_{\tilde w} = 10^{26}$\,sec 
with the branching fractions given in Eq.\,(\ref{eq:branching}).%
\footnote{
In Ref.\,\cite{Ibarra:2008jk},
a similar spectrum is obtained for $\tau_{\tilde w}=2.1\times 10^{26}$\,sec.
The difference of the chosen lifetimes reflects the difference 
of the branching ratio of the $\tilde{w}\to \mu^{\pm}W^\mp$ mode,
which requires the twice larger decay rate in our case.
}
The analysis on the propagation of the electron/positron fluxes in the galaxy is based on 
that given in Ref.\,\cite{Ibarra:2008qg}, and we used numerical approximated Green function
with the choice of the MED propagation model in the reference.
As for the background electron/positron spectra, we used the ones given in the same reference.
The figure shows that the model fits the data of the Fermi experiment quite well.
In the right panel of Fig.\,\ref{fig:electron}, we also show the predicted positron fraction.
In the figure, we have taken into account the solar modulation effect in the 
current solar cycle\,\cite{Baltz:1998xv}.
The figure also show that the positron fraction can be well fitted by the current model.

\begin{figure}[t]
 \begin{minipage}{.38\linewidth}
  \includegraphics[width=\linewidth]{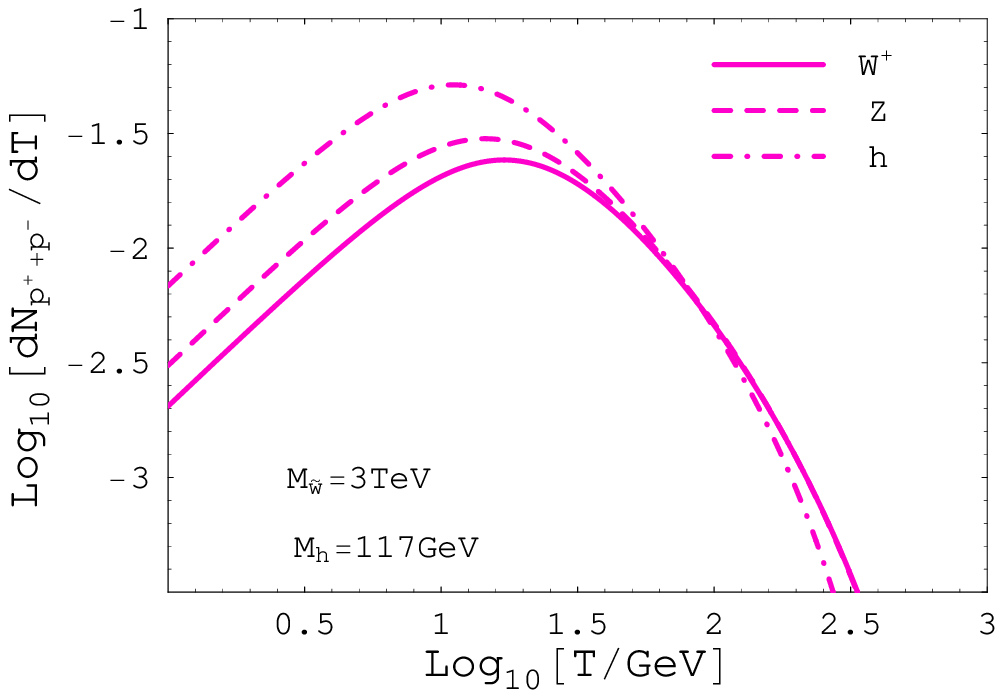}
 \end{minipage}
 \hspace{1cm}
  \begin{minipage}{.34\linewidth}
  \includegraphics[width=1.0\linewidth]{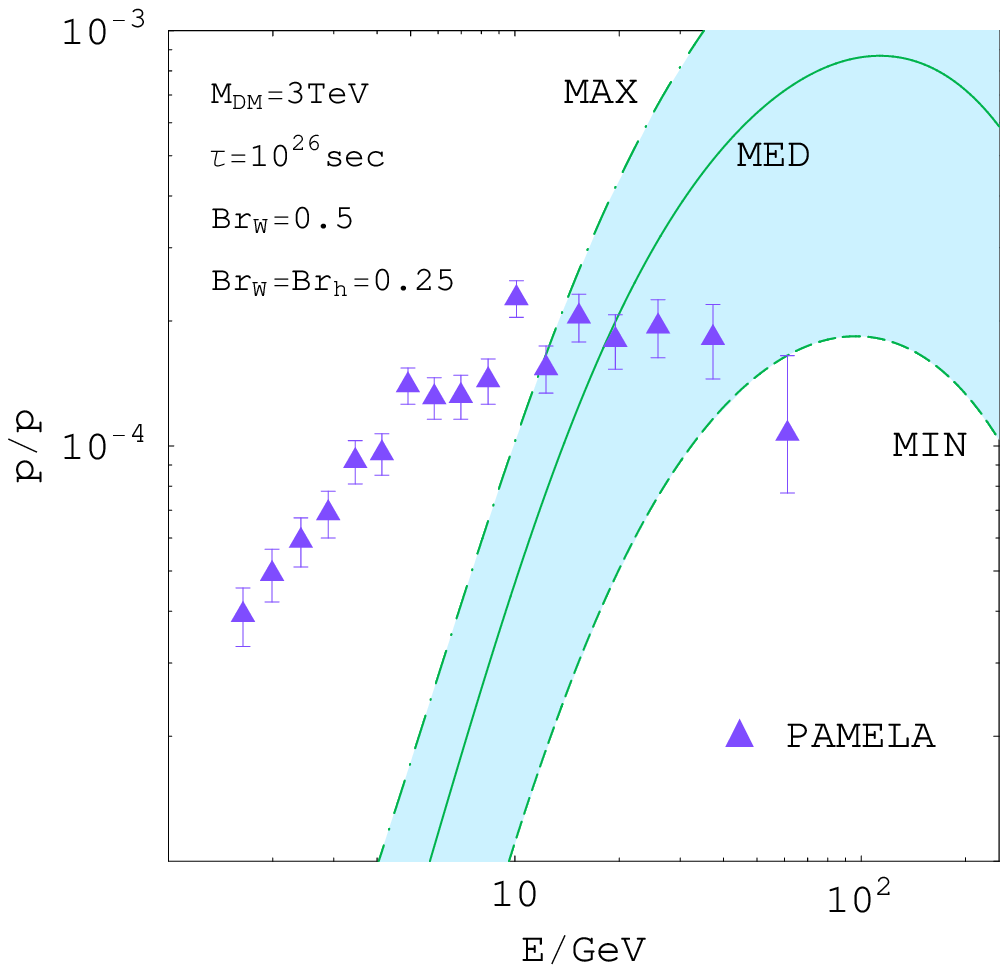}
 \end{minipage}
 \caption{Left) The fragmentation functions of $W$, $Z$ and $h$ into the protons
 and antiprotons.
  We used the program PHYTIA\,\cite{Sjostrand:2006za} to obtain  the functions
  and assumed $m_h = 117$\,GeV.
  Right) Predicted anti-proton and proton ratio in cosmic ray from the dark matter decay
  with the branching ratios in Eq.\,(\ref{eq:branching}).
   The propagations of the proton/antiproton fluxes in the galaxy are analyzed based 
 on a numerically approximated Green function given in Ref.\,\cite{Ibarra:2008qg}
 for three diffusion models, MAX, MED and MIN.
 We compare the prediction with the experimental date in Ref.\,\cite{Adriani:2008zq}.
 The background proton spectrum is borrowed from Ref.\,\cite{Chen:2009gz}.
 }
\label{fig:proton}
\end{figure}

The weak gauge bosons and the higgs boson in the final states of the dark matter
decays also fragment into protons/antiprotons.
Such fragmentations into protons/antiprotons are severely constrained by the PAMELA
experiment which shows no excess in the antiproton fraction\,\cite{Adriani:2008zq}.%
\footnote{
This is a remarkable difference of the non-supersymmetric wino-like dark matter
in comparison with the wino-like dark matter in the supersymmetric context.
In the supersymmetric models such as Ref.\,\cite{Shirai:2009fq}, 
the decay of the wino-like dark matter proceeds via 
the dimension six operators $\tilde w \bar{e}_R\ell_L\ell_L$ which dominate over the 
dimension four operators $\tilde{w}\ell_L H^*$.
On the other hand, in the case of the non-supersymmetric wino-like fermion,
the later dimension four operators are generated radiatively from the former dimension six
operators and the decay process is dominated by the dimension four operators.
This difference may allow us to investigate whether the supersymmetry is behind the wino-like
dark matter  through the observation of the antiproton flux in cosmic ray.
}
In the left panel of Fig.~\ref{fig:proton}, we show the fragmentation functions
into the protons and antiprotons of $W^\pm$, $Z$ and $h$ in the final states of the dark matter decay for $M_{\tilde w}=3$\,TeV.
The figure shows that sizable numbers of the protons/antiprotons are expected from
the fragmentations of those bosons.
In the right panel of Fig.~\ref{fig:proton}, we show the predicted antiproton fraction
in cosmic ray for $\tau_{\tilde w}=10^{26}$\,sec and the branching fraction in Eq.\,(\ref{eq:branching}).
In our analysis, we again used the numerical Green functions of the proton/antiproton propagation
for three different diffusion models given in Ref.\,\cite{Ibarra:2008qg}.
 The background proton spectrum is borrowed from Ref.\,\cite{Chen:2009gz}.
The figure shows that the predicted fraction is contradict with the observed fraction
in some diffusion parameters.
Thus, the decay mode in Eq.\,(\ref{eq:branching}) with $\tau_{\tilde w}=10^{26}$\,sec
for $M_{\tilde w}=3$\,TeV is somewhat disfavored from the antiproton fraction
observed in PAMELA experiments, although not completely
excluded.

Before closing this section, we comment on the cosmic ray from the 
annihilation of the wino-like dark matter.
As discussed in Ref.\,\cite{Hisano:2005ec}, the annihilation cross 
section of the dark matter in our galaxy is also enhanced by the Sommerfeld
enhancement.
The enhancement is, however, not so significant for 2.7\,TeV$\lesssim M_{\tilde w}\lesssim$3\,TeV, 
and the resultant cosmic ray from the annihilation are much smaller
than that from the decay of the dark matter with a lifetime
in the range of $10^{26}$\,sec.

\section{Conclusions}\label{sec:conclusion}
In this study, we found that the small extension of the standard model
with adjoint fermions 
allows the better unification of the three gauge coupling constants
of the standard model with a long enough proton lifetime, 
when the adjoint fermions have masses in ranges of $M_{\tilde w}\lesssim 10^4$\,GeV
and $M_{\tilde g}\lesssim 10^{12}$\,GeV.
We also discussed that the neutral wino-like fermion can be a good candidate for the 
dark matter whose thermal relic density  naturally explains the
observed dark matter density.
With an appropriate choice of the Peccei-Quinn charges, we also found
that the lifetime of the neutral component of the wino-like fermion 
can be an appropriate range to explain the excesses of the electron/positron 
fluxes in cosmic ray in recent experiments.

It should be noted that  the unification scale which is consistent with the dark matter 
density is not much higher than $10^{15}$\,GeV, and hence,
the masses of the heavy gauge bosons which mediate the proton
decay are expected to be close to the current limit, $M_{V}\gtrsim 10^{15.4}$\,GeV.
Therefore, the model predicts rather short lifetime of the proton, $\tau_p =O(10^{34-35})$\,sec, 
which will hopefully be soon detected even at the current detectors such as Super-Kamiokande.
This is a distinctive prediction in comparison with the supersymmetric standard model
where the typical lifetime of the proton decaying via the gauge boson exchange
is $O(10^{36})$\,sec.%
\footnote{In some classes of the grand unified model in the supersymmetric model,
the lifetime of the proton decaying via the gauge boson exchange is rather 
short\,\cite{Ibe:2003ys}. }

The direct searches of the dark matter to detect the recoils of nuclei 
by the dark matter-neclei collision will give a clear evidence of the dark matter.
As shown in Ref.\,\cite{Cirelli:2009uv}, the cross-section of the direct detection
of the wino-like dark matter is around $10^{-45}$\,cm$^{2}$ for $M_{\tilde w}\simeq 3$\,TeV, 
which is within reach of future experiments such as SuperCDMS experiment\,\cite{Brink:2005ej}.

The detection of the wino-like fermion at the collider experiments is also interesting.
As we have mentioned, the charged components of the wino-like fermion decay into 
a neutral wino-like fermion and the charged pion
with the lifetime of ${\cal O}(10^{-10})$\,sec.  
Thus, once they are produced at  collider experiments,
they may leave displaced vertices which help us
to detect the wino-like fermions at future experiments.%
\footnote{See related works for the detection of the wino-like fermion
in the lower mass region at the LHC experiments\,\cite{Ibe:2006de}. }

Finally, we comment on the fate of the gluino-like fermion.
Since it has a rather heavy mass, $M_{\tilde g}\simeq 10^{9-12}$\,GeV,
the cosmic abundance of the gluino-like fermion is highly suppressed 
as long as the temperature of the universe after inflation is much lower
than $M_{\tilde g}$.

\section*{Acknowledgements}
M.~I. appreciate T.T~Yanagida for stimulating discussions.
M.~I. also appreciate S.~Shirai and F.~Takahashi for a lot of advices
in the analysis of cosmic ray spectra.
The work of M.~I. was supported by the U.S. Department of Energy under
contract number DE-AC02-76SF00515.


\begin{thebibliography}{99}  
\bibitem{Langacker:1980js}
  H.~Georgi and S.~L.~Glashow,
  Phys.\ Rev.\ Lett.\  {\bf 32}, 438 (1974);
  H.~Georgi, H.~R.~Quinn and S.~Weinberg,
  Phys.\ Rev.\ Lett.\  {\bf 33}, 451 (1974);
see also the review by,  P.~Langacker,
  Phys.\ Rept.\  {\bf 72}, 185 (1981).
  
  
\bibitem{Ellis:1990wk}
  J.~R.~Ellis, S.~Kelley and D.~V.~Nanopoulos,
  Phys.\ Lett.\  B {\bf 260}, 131 (1991).
  U.~Amaldi, W.~de Boer and H.~Furstenau,
  Phys.\ Lett.\  B {\bf 260}, 447 (1991);
  P.~Langacker and M.~x.~Luo,
  Phys.\ Rev.\  D {\bf 44}, 817 (1991);
  C.~Giunti, C.~W.~Kim and U.~W.~Lee,
  Mod.\ Phys.\ Lett.\  A {\bf 6} (1991) 1745.
\bibitem{:2009gd}
  H.~Nishino {\it et al.}  [Super-Kamiokande Collaboration],
  Phys.\ Rev.\ Lett.\  {\bf 102}, 141801 (2009)
  [arXiv:0903.0676 [hep-ex]].
  
\bibitem{Dimopoulos:1981yj}
  S.~Dimopoulos, S.~Raby and F.~Wilczek,
  Phys.\ Rev.\  D {\bf 24}, 1681 (1981);
  S.~Dimopoulos and H.~Georgi,
  Nucl.\ Phys.\  B {\bf 193}, 150 (1981);
  L.~E.~Ibanez and G.~G.~Ross,
  Phys.\ Lett.\  B {\bf 105}, 439 (1981);
  N.~Sakai,
  Z.\ Phys.\  C {\bf 11}, 153 (1981);
  M.~B.~Einhorn and D.~R.~T.~Jones,
  Nucl.\ Phys.\  B {\bf 196}, 475 (1982);
  W.~J.~Marciano and G.~Senjanovic,
  Phys.\ Rev.\  D {\bf 25}, 3092 (1982).

\bibitem{Krasnikov:1993sc}
  N.~V.~Krasnikov,
  Phys.\ Lett.\  B {\bf 306}, 283 (1993).


\bibitem{Peccei:1977hh}
  R.~D.~Peccei and H.~R.~Quinn,
  Phys.\ Rev.\ Lett.\  {\bf 38}, 1440 (1977);
  R.~D.~Peccei and H.~R.~Quinn,
  Phys.\ Rev.\  D {\bf 16}, 1791 (1977).

\bibitem{Komatsu:2008hk}
  E.~Komatsu {\it et al.}  [WMAP Collaboration],
  Astrophys.\ J.\ Suppl.\  {\bf 180}, 330 (2009)
  [arXiv:0803.0547 [astro-ph]].

\bibitem{Adriani:2008zr}
  O.~Adriani {\it et al.}  [PAMELA Collaboration],
  Nature {\bf 458}, 607 (2009)
  [arXiv:0810.4995 [astro-ph]].

 
\bibitem{Abdo:2009zk}
  A.~A.~Abdo {\it et al.}  [The Fermi LAT Collaboration],
  arXiv:0905.0025 [astro-ph.HE].


  
\bibitem{Bagger:1995bw}
  J.~Bagger, K.~T.~Matchev and D.~Pierce,
  Phys.\ Lett.\  B {\bf 348}, 443 (1995)
  [arXiv:hep-ph/9501277];
  D.~M.~Pierce, J.~A.~Bagger, K.~T.~Matchev and R.~j.~Zhang,
  Nucl.\ Phys.\  B {\bf 491}, 3 (1997)
  [arXiv:hep-ph/9606211].
\bibitem{Ellis:1980jm}
  J.~R.~Ellis, M.~K.~Gaillard, D.~V.~Nanopoulos and S.~Rudaz,
  Nucl.\ Phys.\  B {\bf 176}, 61 (1980).
\bibitem{Aoki:2006ib}
  Y.~Aoki, C.~Dawson, J.~Noaki and A.~Soni,
  Phys.\ Rev.\  D {\bf 75}, 014507 (2007)
  [arXiv:hep-lat/0607002].

  
   
\bibitem{Kim:1979if}
  J.~E.~Kim,
  Phys.\ Rev.\ Lett.\  {\bf 43}, 103 (1979);
  M.~A.~Shifman, A.~I.~Vainshtein and V.~I.~Zakharov,
  Nucl.\ Phys.\  B {\bf 166}, 493 (1980).
\bibitem{Zhitnitsky:1980tq}
  A.~R.~Zhitnitsky,
  Sov.\ J.\ Nucl.\ Phys.\  {\bf 31} (1980) 260
  [Yad.\ Fiz.\  {\bf 31} (1980) 497];
  M.~Dine, W.~Fischler and M.~Srednicki,
  Phys.\ Lett.\  B {\bf 104}, 199 (1981).
  

\bibitem{Baker:2006ts}
  C.~A.~Baker {\it et al.},
  Phys.\ Rev.\ Lett.\  {\bf 97}, 131801 (2006)
  [arXiv:hep-ex/0602020].
  
\bibitem{Raffelt:2006cw}
  G.~G.~Raffelt,
  Lect.\ Notes Phys.\  {\bf 741}, 51 (2008)
  [arXiv:hep-ph/0611350].
\bibitem{Nardi:2008ix}
  E.~Nardi, F.~Sannino and A.~Strumia,
  JCAP {\bf 0901}, 043 (2009)
  [arXiv:0811.4153 [hep-ph]];
  P.~Meade, M.~Papucci, A.~Strumia and T.~Volansky,
  arXiv:0905.0480 [hep-ph].
  
\bibitem{Ibarra:2008jk}
  A.~Ibarra and D.~Tran,
  JCAP {\bf 0902}, 021 (2009)
  [arXiv:0811.1555 [hep-ph]];
  A.~Ibarra, D.~Tran and C.~Weniger,
  arXiv:0906.1571 [hep-ph].

\bibitem{Shirai:2009fq}
  S.~Shirai, F.~Takahashi and T.~T.~Yanagida,
  arXiv:0905.0388 [hep-ph].
\bibitem{Chen:2009mj}
  C.~H.~Chen, C.~Q.~Geng and D.~V.~Zhuridov,
  arXiv:0905.0652 [hep-ph].
\bibitem{Feng:1999fu}
  J.~L.~Feng, T.~Moroi, L.~Randall, M.~Strassler and S.~f.~Su,
  Phys.\ Rev.\ Lett.\  {\bf 83}, 1731 (1999).
  
\bibitem{Fujii:2003nr}
See, for example, the result in  M.~Fujii, M.~Ibe and T.~Yanagida,
  Phys.\ Lett.\  B {\bf 579}, 6 (2004)
  [arXiv:hep-ph/0310142].
  
\bibitem{Hisano:2006nn}
  J.~Hisano, S.~Matsumoto, M.~Nagai, O.~Saito and M.~Senami,
  Phys.\ Lett.\  B {\bf 646}, 34 (2007)
  [arXiv:hep-ph/0610249].

\bibitem{Ishiwata:2009vx}
  K.~Ishiwata, S.~Matsumoto and T.~Moroi,
  JHEP {\bf 0905}, 110 (2009)
  [arXiv:0903.0242 [hep-ph]].

\bibitem{Buchmuller:2009xv}
  W.~Buchmuller, A.~Ibarra, T.~Shindou, F.~Takayama and D.~Tran,
  arXiv:0906.1187 [hep-ph].

\bibitem{Sjostrand:2006za}
  T.~Sjostrand, S.~Mrenna and P.~Skands,
  JHEP {\bf 0605}, 026 (2006)
  [arXiv:hep-ph/0603175].
  
\bibitem{Ibarra:2008qg}
  A.~Ibarra and D.~Tran,
  JCAP {\bf 0807}, 002 (2008)
  [arXiv:0804.4596 [astro-ph]].
  


\bibitem{Chang:2008zz}
  J.~Chang {\it et al.},
  Nature {\bf 456}, 362 (2008).
\bibitem{Torii:2008xu}
  S.~Torii {\it et al.},
  arXiv:0809.0760 [astro-ph].
\bibitem{Collaboration:2008aaa}
  F.~Aharonian {\it et al.}  [H.E.S.S. Collaboration],
  Phys.\ Rev.\ Lett.\  {\bf 101}, 261104 (2008)
  [arXiv:0811.3894 [astro-ph]].



\bibitem{Baltz:1998xv}
  E.~A.~Baltz and J.~Edsjo,
  Phys.\ Rev.\  D {\bf 59} (1999) 023511
  [arXiv:astro-ph/9808243].
\bibitem{Aguilar:2007yf}
  M.~Aguilar {\it et al.}  [AMS-01 Collaboration],
  Phys.\ Lett.\  B {\bf 646}, 145 (2007)
  [arXiv:astro-ph/0703154].

\bibitem{Barwick:1997ig}
  S.~W.~Barwick {\it et al.}  [HEAT Collaboration],
  Astrophys.\ J.\  {\bf 482}, L191 (1997)
  [arXiv:astro-ph/9703192].


\bibitem{Adriani:2008zq}
  O.~Adriani {\it et al.},
  Phys.\ Rev.\ Lett.\  {\bf 102}, 051101 (2009)
  [arXiv:0810.4994 [astro-ph]].
  

\bibitem{Chen:2009gz}
  C.~R.~Chen, M.~M.~Nojiri, S.~C.~Park, J.~Shu and M.~Takeuchi,
  arXiv:0903.1971 [hep-ph].


\bibitem{Hisano:2005ec}
  J.~Hisano, S.~Matsumoto, O.~Saito and M.~Senami,
  Phys.\ Rev.\  D {\bf 73}, 055004 (2006)
  [arXiv:hep-ph/0511118].
 
\bibitem{Ibe:2003ys}
  M.~Ibe and T.~Watari,
  Phys.\ Rev.\  D {\bf 67}, 114021 (2003)
  [arXiv:hep-ph/0303123].
  
\bibitem{Cirelli:2009uv}
  M.~Cirelli and A.~Strumia,
  arXiv:0903.3381 [hep-ph].
\bibitem{Brink:2005ej}
  P.~L.~Brink {\it et al.}  [CDMS-II Collaboration],
{\it In the Proceedings of 22nd Texas Symposium on Relativistic Astrophysics at Stanford University, Stanford, California, 13-17 Dec 2004, pp
2529}
  [arXiv:astro-ph/0503583].
  

\bibitem{Ibe:2006de}
  M.~Ibe, T.~Moroi and T.~T.~Yanagida,
  Phys.\ Lett.\  B {\bf 644}, 355 (2007)
  [arXiv:hep-ph/0610277].
 


\end{thebibliography}
\end{document}